\documentclass[%
 aip,
 apl,
 amsmath,amssymb,
 numerical,
 reprint,%
]{revtex4-2}
\usepackage{booktabs,multirow,array}
\usepackage[table]{xcolor} 
\usepackage{graphicx}
\usepackage{dcolumn}
\usepackage{bm}
\usepackage{hyperref}
\usepackage{amsmath}

\usepackage{comment}
\usepackage[utf8]{inputenc}
\usepackage[T1]{fontenc}
\usepackage{mathptmx}
\usepackage{siunitx}
\usepackage{booktabs}
\usepackage{tabularx}

\usepackage{xcolor}
\usepackage{siunitx}
\usepackage{appendix}
\usepackage{subfig}
\usepackage{todonotes}
\makeatletter
\renewcommand{\todo}[2][]{%
    \@ifnextchar\bgroup{%
        \@todo[size=\footnotesize,#1]{#2}%
    }{%
        \@todo[size=\footnotesize,#1]{#2}%
    }%
}
\makeatother
\usepackage[labelfont=bf,
   justification=justified,
   format=plain]{caption} 
\usepackage{color,soul}
\graphicspath{{figures/}}

\begin{document}

\preprint{AIP/123-QED}

\title[Heliometric stereo: a new frontier in surface profilometry]{Heliometric stereo: a new frontier in surface profilometry}


\author{\underline{Aleksandar Radi\'{c}}}
\email[Author to whom correspondence should be addressed: ]{ar2071@cam.ac.uk}
\affiliation{Department of Physics, Cavendish Laboratory, 19 J.J. Thomson Avenue, University of Cambridge, Cambridge, CB3 0HE, UK}

\author{\underline{Sam M. Lambrick}}
\affiliation{Department of Physics, Cavendish Laboratory, 19 J.J. Thomson Avenue, University of Cambridge, Cambridge, CB3 0HE, UK}
\affiliation{ 
Ionoptika Ltd, Units B5-B6, Millbrook Close, Chandlers Ford, Southampton, S053 4BZ, UK}

\author{Chenyang Zhao}
\affiliation{Department of Physics, Cavendish Laboratory, 19 J.J. Thomson Avenue, University of Cambridge, Cambridge, CB3 0HE, UK}

\author{Nick A. von Jeinsen}
\affiliation{Department of Physics, Cavendish Laboratory, 19 J.J. Thomson Avenue, University of Cambridge, Cambridge, CB3 0HE, UK}

\author{Andrew P. Jardine}
\affiliation{Department of Physics, Cavendish Laboratory, 19 J.J. Thomson Avenue, University of Cambridge, Cambridge, CB3 0HE, UK}

\author{David J. Ward}
\affiliation{Department of Physics, Cavendish Laboratory, 19 J.J. Thomson Avenue, University of Cambridge, Cambridge, CB3 0HE, UK}
\affiliation{ 
Ionoptika Ltd, Units B5-B6, Millbrook Close, Chandlers Ford, Southampton, S053 4BZ, UK}

\author{Paul C. Dastoor}
\affiliation{Department of Physics, Cavendish Laboratory, 19 J.J. Thomson Avenue, University of Cambridge, Cambridge, CB3 0HE, UK}
\affiliation{Centre for Organic Electronics, Physics Building, University of Newcastle, Callaghan, NSW 2308, Australia}
\date{\today}

\begin{abstract}
\noindent Accurate and reliable measurements of three-dimensional surface structures are important for a broad range of technological and research applications, including materials science, nanotechnology, and biomedical research. Scanning helium microscopy (SHeM) uses low-energy ($\sim\SI{64}{\milli eV}$) neutral helium atoms as the imaging probe particles, providing a highly sensitive and delicate approach to measuring surface topography. To date, topographic SHeM measurements have been largely qualitative, but with the advent of the heliometric stereo method –- a technique that combines multiple images to create a 3D representation of a surface –- quantitative maps of surface topography may now be acquired with SHeM. Here, we present and discuss two different implementations of heliometric stereo on two separate instruments, a single detector SHeM and a multiple-detector SHeM. Both implementations show good accuracy (5\% and 10\% respectively) for recovering the shape of a surface. Additionally, we discuss where heliometric stereo is most applicable, identify contrast features that can limit its accuracy, and discuss how to mitigate these limitations with careful design and sample choices that be readily implemented on current instruments.
\end{abstract}

\maketitle
\section{Introduction}
\noindent Accurate measurements of surface topography (often termed surface metrology) are crucial to modern research and development, with widespread applications from understanding material behavior to optimizing manufacturing\cite{jiang_paradigm_2007,jiang_paradigm_2007-1}.
Significantly, surface metrology can also provide assurance that products and materials meet required standards for functionality, quality, and safety\cite{Mathia2011}. As new technologies emerge, for example, in micro/nanoscale organic and quantum devices, reliable measurement of delicate surface topographies on an ever smaller scale is crucial. Scanning helium microscopy (SHeM)\cite{Witham2011,Barr2014,Koch2008,palau_neutral_2023,flatabø2023reflection} is a technique that uses a narrow beam of neutral helium atoms ($\geq\SI{300}{nm}$), generated \textit{via} a collimating pinhole to produce topographic micrographs. The use of a neutral atom beam has two key advantages for topographic measurements. First, SHeM uses a low energy beam, with a typical de Broglie energy and wavelength of  $\SI{64}{meV}$ and $\SI{0.06}{nm}$, respectively \cite{Holst2021,lifPaper}. Thus rendering beam damage, as observed with charged beam methods\cite{Ramachandra2009}, impossible. Second, thermal helium atoms scatter off the outermost electron density distribution which results in information exclusively from the surface, without any contribution due to beam penetration into the bulk\cite{Holst2021}. The atomic scale wavelength of the helium atoms gives rise to cosine-like scattering from almost any surface\cite{LambrickDiffuse2022}, with the exception being atomically flat and adsorbate-free crystals\cite{lifPaper,hatchwell2023measuring}, giving the technique a distinct \emph{lack} of non-topographic contrast mechanisms that place it ideally for profilometry of most any surface. Existing non-contact techniques, such as optical profilometry, laser scanning confocal microscopy, scanning electron microscopy, interferometric profilometry and structured light scanning\cite{Kournetas2017,Arvidsson2006,Valverde2013}, either use highly energetic probe particles, which can induce changes in the sample, or can have a non-negligible interaction volume with the surface and thus do not measure the surface topography exclusively. These methods also have a variety of secondary contrast mechanisms that can complicate topographic reconstruction whereas SHeM almost exclusively exhibits diffuse scattering. Contact techniques, such as stylus profilometers, scanning tunneling microscopes and atomic force microscopes\cite{Al-Nawas2001,Rupp2003,Svanborg2010,Wennerberg2014}, provide a more direct surface profile measurement through contact with a physical probe. However, these are largely limited in either lateral resolution (stylus profilometers) or in aspect ratio (AFM/STM), can induce changes in the sample due to the direct contact with the probe, and in all cases convolution with the probe tip must be considered.  

SHeM has demonstrated its ability to measure surface topography qualitatively\cite{Fahy2018,Lambrick2018,fahy_highly_2015,pranav}, and two previous publications have shown SHeM to be capable of providing individual pieces of 3D information. Myles et al.\cite{Myles2019} applied stereophotogrammetry to taxonomy, measuring specific dimensions of biological specimens. Lambrick et al.\cite{LambrickMultiple2020} demonstrated that quantitative 3D measurements can be made by interpreting specific contrast features of SHeM to depth profile trenches in silicon precisely. Both approaches make accurate measurements of specific dimensions, but do not yield a 3D view of the entire sample. 

In this paper, we describe the development of 3D imaging using SHeM (a technique termed heliometric stereo) using both single detector \cite{Radic_3d_2024}, and multiple detector instruments\cite{chenyang_bshem}. We provide detailed analysis of heliometric stereo reconstruction accuracy, how it is enabled and affected by contrast mechanisms unique to SHeM, and how these sources of error can be mitigated using instrumentation, algorithms and/or careful sample selection\cite{Lambrick3d}. Sections \ref{sec:heliometric_stereo}, \ref{sec:theory} and \ref{sec:single-detector} restate the heliometric stereo method conceptually, the supporting theory and first experimental surface reconstructions using a single detector, respectively, as presented by Radi\'{c} et al.\cite{Radic_3d_2024} and Lambrick \& Palau et al.\cite{Lambrick3d}. The version presented in the current work extends the original analysis and discussion than was previously presented by Radi\'{c} et al.\cite{Radic_3d_2024}. In sections \ref{sec:heliometric_stereo}-\ref{sec:single-detector} we include additional discussion and analysis on the original results that lies beyond the scope of the original manuscripts' Letter formats, providing a complete, and in-depth, understanding of heliometric stereo such that this manuscript may be an entirely self-contained reference of the method in both theory and experiment. Previously published content is indicated in text with the relevant citation. Sections \ref{sec:multi_detector}-\ref{sec:masking_shadowing} present previously published results of multi-detector reconstructions, by Zhao et al.\cite{chenyang_bshem}, alongside new data in order to provide a comprehensive error analysis on heliometric stereo, highlighting under what circumstances the method can be applied with high accuracy.

\section{Heliometric Stereo}\label{sec:heliometric_stereo}
Analogously to photometric stereo in optics, heliometric stereo is based on the assumption that a surface will scatter the incident particles/waves with a known scattering function, formally termed the bidirectional reflection distribution function (BRDF)\cite{StoverOptics}. The BRDF defines the scattering probability as a function of the incident direction and outgoing scattered direction of the helium atoms relative to the surface orientation. In a SHeM, if the microscope detector is fixed at a known position and angle relative to the surface and the direction of the incident helium beam is known, then the remaining variable in the BRDF describing the recorded intensity is the surface orientation. The well defined scattering geometry of SHeM\cite{LambrickDiffuse2022} satisfies the above description. Therefore to perform 3D surface mapping all that is needed in addition to the recorded intensity (helium micrographs) is a description of the BRDF. 

Recent experimental results have shown that the majority of unprepared surfaces studied in SHeM exhibit almost exclusively diffuse, or close to diffuse, scattering \cite{LambrickDiffuse2022,LambrickMultiple2020,Lambrick2018,Fahy2018,hatchwell2023measuring,RadicMPhil,witham_increased_2012,Witham2014}. The observation of diffuse scattering provides the model for the BRDF (generally referred to simply as the scattering distribution in SHeM) required to perform surface reconstructions.  Crucially, these observations enabled the first experimental implementation of heliometric stereo\cite{Radic_3d_2024}. These materials have typically been unprepared `technological samples', in contrast to conventional surface science studies where surfaces are prepared and maintained in a pristine, atomically perfect state, and where specular or Bragg scattering is often dominant\cite{Holst2021,lifPaper,estermann_beugung_1930,radic_defects,radic_2d_methods,corem_ordered_2013}. The observation of diffuse scattering from technological surfaces arises from the fact that the de Broglie wavelength of the helium atoms ($\lambda\sim\SI{1}{\angstrom}$) is comparable to the interatomic spacing in solids. As such, given that most technological samples have atomically rough surfaces, diffuse scattering dominates. Consequently, the observed diffuse distribution is consistent with Knudsen's cosine law\cite{Knudsen1935}, and is analogous to Lambert's cosine law for visible light\cite{Lambert1760}. 

It is important to note that for SHeM there are known exceptions to the model of diffuse scattering used in heliometric stereo, the primary exception being multiple scattering, a contribution to the detected signal that arises from more than one scattering event occurring between the source and detector apertures. Discussion on the effects of multiple scattering on contrast, and therefore reconstruction accuracy, is presented theoretically in section \ref{sec:theory}, and experimentally in section \ref{sec:masking_shadowing}. 

\begin{figure}
    \centering
    \small{\textsf{\input{figures/heliometric_stereo_construction_flip.tex}}}
    \caption{The mechanism of image formation with (a) ordinary optical imaging and (b) SHeM. With light, a broad illumination of the surface occurs, with the scattered light being focused by a camera. In SHeM it is the incident beam that is focused onto the surface, with the atoms scattered in particular directions detected. In photometric stereo with light there is a relationship between the direction of illumination, $\hat{\bm{n}}_i$, and the surface normal, in SHeM the relationship is between the direction of detection and the surface normal. Figure adapted from Lambrick \& Salvador Palau et al.\cite{Lambrick3d}. Reproduced with permission from \textit{Phys. Rev. A} 103, 053315 (2021). Copyright American Physical Society.}
    \label{fig:light_shem_equivalence}
\end{figure}

\section{Theory}\label{sec:theory}

Here we restate the theoretical basis of heliometric stereo as first presented by Lambrick \& Palau et al.\cite{Lambrick3d} with text and figures largely unchanged from the original manuscript so that the current work may serve as a complete reference for the theory and experimental implementations of heliometric stereo to-date. The foundation of the method allows us to build up the two experimental implementations in sections IV and V, while the description below is self-contained for the purposes of understanding the following experimental results, full details, including supporting calculations, can be found in Lambrick \& Palau\cite{Lambrick3d}.

Mathematically, diffuse scattering follows a cosine distribution centered on the surface normal,
\begin{equation}\label{eq:cosine}
    I(\theta) = \rho\cos\theta = \rho\hat{\bm{n}}\cdot\hat{\bm{d}},
\end{equation}
where $\theta$ is the angle between the local surface normal, $\hat{\bm{n}}$, and the outgoing scattering direction $\hat{\bm{d}}$. It can be shown that for cosine scattering with a macroscopic circular detection area -- such as is present experimentally in SHeM\cite{Barr2014,chenyang_bshem} -- that equation \eqref{eq:cosine} holds the same mathematical form up to a multiplicative constant\cite{Lambrick3d}. Equation \eqref{eq:cosine} has the same form for light in photometric stereo\cite{woodham_photometric_1980}, however, due to differences in image projection for light, $\hat{\bm{d}}$ represents the illumination direction (the outgoing scattering direction). The equivalence between detection and illumination in SHeM and in ordinary optical imaging are shown in figure \ref{fig:light_shem_equivalence}. With light, there is a broad illumination from a light source and the scattered light is focused to form an image in the camera. In SHeM a focused beam is produced that scatters from the sample with a selection of the scattered atoms detected, the focused beam is then scanned over the sample to create an image \emph{via} Helmholtz reciprocity. With light, Lambertian cosine scattering\cite{Lambert1760} places a cosine dependence on the angle between the illumination direction and the surface normal, which leads to the same equation \eqref{eq:cosine}, just with a different definition of $\hat{\bm{d}}$. Both heliometric- and photometric-stereo then follow the same mathematical procedure to produce a topographic reconstruction.

As equation \eqref{eq:cosine} directly relates the intensity scattered in a particular direction to the orientation of the surface, it is possible to acquire the surface orientation from a series of measured intensities. In SHeM that means multiple different detection directions, $\hat{\bm{d}}_i$ for the same illumination direction, see an example in figure \ref{fig:light_shem_equivalence}; for photometric stereo with light that would mean multiple separate illumination directions with the same camera position. 

For multiple detection directions in a SHeM system, equation \eqref{eq:cosine} may be written in matrix form,
\begin{equation}\label{eq:ch3D:basic_photostereo}
	\Vec{I}_{(x',y')}=\rho \mathsf{D}\hat{\bm{n}},
\end{equation}
where $\Vec{I}$ is a \(m\)-dimensional vector of pixel intensities, all for the same position on the sample, corresponding to \(m\) images taken with different detection directions. $\mathsf{D}$ is a \(m \times 3\) matrix containing the normalized vectors pointing from the scattering point to the detectors. The linear system is thus solved,
\begin{gather}
	\rho_{(x^\prime,y^\prime)} = |\mathsf{D}^{-1}\Vec{I}_{(x^\prime,y^\prime)}|,\label{eq:ch3D:solvePhotostereo1}\\
	\hat{\bm{n}}_{(x^\prime,y^\prime)} = \frac{1}{\rho_{(x^\prime,y^\prime)}}\mathsf{D}^{-1}\Vec{I}_{(x^\prime,y^\prime)}.\label{eq:ch3D:solvePhotostereo2}
\end{gather}
Assuming the height of the surface can be described by a continuous function of the lateral position, {\em i.e.} $z = f(x, y)$, then
\begin{gather}
	\hat{\bm{n}}(x,y) = \bm{\nabla} F(x,y,z) = \bm{\nabla}[z- f(x,y)].\label{eq:ch3D:gradient_field}
\end{gather}
Thus once the surface normals are found, the gradient field given by equation \eqref{eq:ch3D:gradient_field} may be integrated to obtain an equation $z=f(x,y)$ of the surface profile, \emph{i.e.} a topographic map of the sample. A regularized least squares approach, developed by Harker and O'leary\cite{HarkerOLeary2008,HarkerMatlab}, is used in the current work. 

The orientation of a surface can be defined by two angles, plus, following the formalism of photometric stereo, we allow for a constant of proportionality, the albedo factor ($\rho$ in equation \eqref{eq:cosine}), hence there are 3 degrees of freedom. The albedo factor is borrowed from light scattering and has the same meaning in the current context: the albedo factor is the proportion of scattered flux that is diffusely scattered, with the rest of the flux not following equation \ref{eq:cosine}, and can therefore vary across a sample due to roughness or material, or contrast features such as masking and multiple scattering. The albedo factor, along with surface normal, is calculated per-pixel when solving the system of linear equations in \ref{eq:ch3D:gradient_field}. Therefore, a minimum of 3 data points (detector directions) are needed for each position on the surface to reconstruct the 3D topography. In practice SHeM data is often signal-to-noise  limited and there may be minor deviations to the cosine law, thus at least 4 are recommended, but as demonstrated by our results in section \ref{sec:multi_detector}, not necessarily required, for a reconstruction to be robust.  Indeed, the method has proved robust to noise in simulated SHeM micrographs\cite{Lambrick3d}.

\section{Reconstruction using a single-detector instrument}
\label{sec:single-detector}

In this section we present previous results on the experimental implementation of single detector heliometric stereo by Radi\'{c} et al.\cite{Radic_3d_2024}. The results, and supporting discussion, on the single-detector implementation of heliometric stereo presented in this section is largely unchanged from the original manuscript\cite{Radic_3d_2024}, but included in full detail such that the current manuscript can serve as a complete record of heliometric stereo theory and its experimental implementations. Figures \ref{fig:pinhole_plate}, \ref{fig:BigFigure}, \ref{fig:measurements}, and the corresponding text, have been reproduced with permission from Radi\`{c} et al.\cite{Radic_3d_2024}, with extensions to the analysis and discussion specifically concerning the impacts of masking, shadowing and multiple scattering.

A key result found in previous work\cite{Lambrick3d} was that \emph{rotations about the beam axis} in SHeM may be used to create multiple `effective detection directions', on an instrument with a single detector. In figure \ref{fig:rotating_detectors} that acquisition process is illustrated: an image is taken with the sample in one orientation (a), then the sample is rotated and a second image is taken (b). As the rotation occurred around the beam axis, the rotation of the sample is equivalent to a rotation of the detector (c). Therefore multiple detection directions may be sampled in an instrument with a single physical detector (d). However, if the rotation is performed about an axis other than one parallel to the beam axis the illumination of the sample is changed, and thus the correspondence between sample rotations and multiple detectors is lost \cite{Lambrick3d}.

\begin{figure}
    \centering

    \textsf{\small\input{figures/heliometric_stereo_rotating_sample.tex}}
    
    \caption{Illustration of multiple image acquisition with a single detector instrument; (a) a helium micrograph is obtained. (b) the sample is rotated about the helium beam axis to obtain a second micrograph. (c) the rotation of the sample is equivalent to a rotation of the detector. (d) We now have 2 micrographs with 2 different detection directions, corresponding to 2 effective detectors. The images with multiple detection directions can now be used to perform a heliometric stereo reconstruction. Note that a rotation of the sample about an axis other than the helium beam axis would result in a different illumination of the sample and therefore the correspondence with a rotation of the detector would be lost. Figure adapted with permission from Radi\'{c} et al.\cite{Radic_3d_2024}. Reproduced from \textit{Appl. Phys. Lett.} 124, 204101 (2024), with the permission of AIP Publishing.}
    \label{fig:rotating_detectors}
\end{figure}

\begin{figure}
	\centering
	\textsf{\small \input{figures/new_pinhole_plate_v2_1.tex}}

    \caption{Cross-sectional diagram of the sample manipulation and pinhole plate of the single detector SHeM at Cambridge, ``A-SHeM''. (a) with the rotation stage introduced to the original, $\theta=\ang{45}$, configuration, as used by von Jeinsen et al.\cite{lifPaper}. (b) with the new pinhole plate macroscopically rotating the system to operate in a normal incidence configuration. Highlighted in yellow are the modular pinhole plates, as distinct from the fixed vacuum chamber walls. The new design adds a small, but largely negligible, distance of $\sim\SI{1.5}{\centi\metre}$ to the overall pathway to the ionization volume of the detector, which is $\sim\SI{60}{\centi\metre}$. Figure adapted with permission from Radi\'{c} et al.\cite{Radic_3d_2024}. Reproduced from \textit{Appl. Phys. Lett.} 124, 204101 (2024), with the permission of AIP Publishing.}
	\label{fig:pinhole_plate}
\end{figure}

The single detector SHeM at Cambridge (A-SHeM), was designed to operate with an incidence angle of $\ang{45}$\cite{Barr2014} enabling topographic imaging as well as studies of specular scattering. To perform heliometric-stereo reconstruction it was necessary to the A-SHeM to operate in a normal incidence configuration through a new 3D printed `pinhole plate' optical element\cite{m_bergin_complex_2021,radicPlastics}. The pinhole plate mounts the collimating pinhole, defines the detection geometry and, in the A-SHeM, is used to mount the sample and sample manipulation stages. The old ($\ang{45}$) (a) and new ($\ang{90}$) (b) configurations are shown in figure \ref{fig:pinhole_plate}, with the incident helium beam and rotation axis highlighted. The new pinhole plate macroscopically changes the mounted orientation of both the sample and the sample manipulation stages relative to the incident beam: in figure \ref{fig:pinhole_plate} (b) the beam is now incident at $\ang{0}$ to the sample and the axis of rotation is parallel to the beam. However, while the rotation axis and the beam are parallel, they are not coaxial, thus an alignment procedure was needed to ensure that the reconstructed image area remained centered in the micrographs as the sample was rotated. The same rotation and alignment procedure as used by von Jeinsen et al.\cite{lifPaper} was applied here. While the the new design does introduce a small ($\sim3\%$) change in the path length from the collection aperture to the detector (figure \ref{fig:pinhole_plate}), the detection efficiency, and therefore image acquisition, is unaffected\cite{Lambrick2018}.

\begin{figure*}
    \centering
    \includegraphics[width=\linewidth]{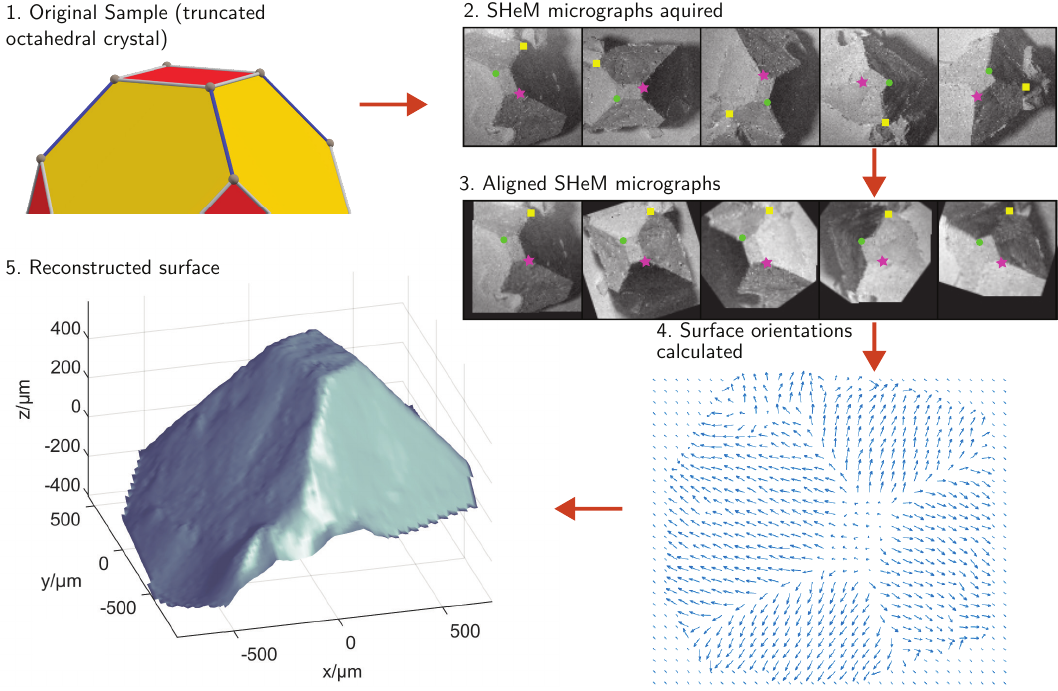}
    \caption{Experimental procedure of performing heliometric stereo. (1)-(2) A series of SHeM micrographs are taken of the sample. (2)-(3) Reference points are used to align the micrographs. (4) the detected helium intensities from the aligned micrographs are used with equation \eqref{eq:ch3D:solvePhotostereo2} to give surface normal vectors. (5) the normal vectors are integrated to a give a surface topography map. Micrographs acquired with $\SI{20}{\micro\metre}$ pixel size. Full imaging details in section SII. Figure reproduced with permission from Radi\'{c} et al. \cite{Radic_3d_2024}.  Reproduced from \textit{Appl. Phys. Lett.} 124, 204101 (2024), with the permission of AIP Publishing.\newline\footnotesize{Panel 1 adapted from a figure by T. Piesk, CC BY 4.0, 2018, \url{https://commons.wikimedia.org/w/index.php?curid=66347083.}}}
    \label{fig:BigFigure}
\end{figure*}

To test the reconstruction accuracy for the A-SHeM implementation, a sample of well-defined topography was needed that matched the basic assumptions of heliometric stereo, principally a continuous $z=f(x,y)$ surface. Aluminum potassium sulfate crystals ($KAl(SO_{4})_{2}. 12 H_{2}O$) were chosen. They possess a well-defined octahedral structure and can be readily grown to form single crystals on micron to millimeter length-scales. To demonstrate topographic reconstruction, we acquired helium micrographs at five equally spaced azimuthal intervals of $\ang{72}$. Over-sampling of the system (equation \eqref{eq:ch3D:solvePhotostereo2}) was used to account for noisy data and/or for slight deviations from ideal diffuse scattering. In addition, the rotational stage of the sample manipulator has a small amount of mechanical inaccuracy, which limits the accuracy to which the center of rotation can be determined to a few pixels. Consequently, the tracking of the sample through rotations was imperfect and thus collecting more micrographs minimized the image area where fewer than three measurements were present, in total 24\% of the image area could not be reconstructed.

The reconstruction process is illustrated in figure \ref{fig:BigFigure} and starts by manually identifying common points across the micrographs to correlate and align the separate azimuthal images. Distinct points on the sample were selected that are accurately identifiable across micrographs, as shown in panel 2.  A Euclidean (or rigid) transformation matrix, one where only translation and rotation are permitted, was calculated to create a mapping of each micrograph onto the same axis. Mapping to a common set of axes is crucial because the pixels in each image must correspond to the same point on the sample surface so that surface orientation may be inferred. Panel 3 shows the images after alignment, with 3 tracking points highlighted (out of a total 6 used for image alignment). Where at least 3 micrographs overlapped, the intensities from the micrographs were used to calculate the surface normals according to equation \eqref{eq:ch3D:solvePhotostereo2}.  Normal vectors for each pixel were generated, as shown in Figure \ref{fig:BigFigure} panel 4. Pixels that cannot be correlated across at least 3 micrographs are assigned a surface orientation parallel to the $z$ axis because otherwise the linear system, shown in equation \eqref{eq:ch3D:basic_photostereo}, becomes underdetermined. Orientation parallel to the $z$ axis was chosen because it is the average of all possible orientations and thus underdetermined pixels have minimal effect on the integration of surface normal vectors which gives the final surface reconstruction.  Finally, the normal map is integrated using the regularized least squares method\cite{HarkerOLeary2008,HarkerMatlab}, giving the surface reconstruction, shown in panel 5. The reconstruction captures all qualitative aspects of the crystal: we are clearly observing the top half of a truncated octahedron.

Two metrics were employed to assess the quantitative accuracy of the reconstruction. First, the surface area of the top of the crystal was determined using 2D SHeM micrographs and independently via SEM (figure S1) to ensure the reconstruction process does not distort the size and shape of the sample. It should be noted that measurements taken directly from 2D SHeM micrographs and SEM images will both be affected by any tilt in the sample, or sample mounting. Second, the angles between the facets of the reconstructed surface were measured and compared to the known facet angles of an octahedral aluminum potassium sulfate crystal.

The dimensions measured for the size of the top face of the truncated pyramid are given in table \ref{tab:pyramid_measurements}. All of the  dimensions agree to within experimental uncertainty, indicating that there is no significant distortion occurring due to the reconstruction algorithm.

\begin{table}[h]
\centering
    \begin{tabular}{@{} l*{4}{>{$}c<{$}} @{}}
    \toprule
     & \multicolumn{4}{c@{}}{Dimensions of Pyramid Truncated Face}\\
    \cmidrule(l){2-5}
     & Long & Short & Area & Units\\
    \midrule
    \multirow{3}{*}{SEM}  &240.5\pm 5 & 137.2\pm5 & \num{3.30e4}\pm 25 & \mathrm{px}\\
    & 438\pm 9 & 250\pm 9 & 110 \pm 0.005 & \si{\micro\metre}  \\
    
    \multirow{3}{*}{}\\
    
    \multirow{3}{*}{2D SHeM}  & 28.2\pm 1 & 16.5\pm 1 & 466 \pm 1 & \mathrm{px}\\ 
     & 423\pm 15 & 248 \pm 15 & 105 \pm 7 & \si{\micro\metre} \\
    
    \multirow{3}{*}{}\\
    
    \multirow{3}{*}{3D SHeM}  & 154.9\pm 1 & 92.6 \pm 1 & \num{1.43e3} \pm 1 & \mathrm{px}\\ 
    & 424 \pm 13 & 254 \pm 13 & 108 \pm 6 & \si{\micro\metre} \\

    \bottomrule
    \end{tabular}
    \caption{Measured side lengths, and calculated areas, of the top face of the truncated pyramid in figures 3 and 4 used to evaluate heliometric stereo reconstruction accuracy. Pixel to millimeter conversions, SEM: $\SI{1000}{\micro\metre} = 549\pm\SI{2}{px}$, 2D SHeM: $\SI{1000}{\micro\metre} = 66.7\pm\SI{1}{px}$, 3D SHeM: $\SI{1000}{\micro\metre} = 365\pm\SI{1}{px}$. SEM image used for measurement shown in figure S1. Table adapted with permission from Radi\'{c} et al.\cite{Radic_3d_2024}.}
    \label{tab:pyramid_measurements}
\end{table}

The angle between the top facet of the crystal and the side facets for a truncated octahedron are $\ang{125}$ (to 3 s.f.). To measure the angle between the reconstructed facets, the surface height going left to right across the crystal and top to bottom were averaged, producing two data sets each crossing the top of a trapezoid. Figure \ref{fig:measurements} shows a surface height map of the reconstruction in figure \ref{fig:BigFigure}, with the extracted profiles horizontally and vertically across the crystal. Linear regression was applied to each edge of the two trapezoidal plots, the regression lines shown in color in figure \ref{fig:measurements}, resulting in four measurements of the facet angle; $\ang{136}$, $\ang{122}$, $\ang{129}$ and $\ang{131}$, with a mean value of $\ang{129}$ a $5\%$ overestimate compared to the ideal octahedron. The uncertainties in the linear regression were small, therefore the deviations from the expected value are not attributed to random noise.

\begin{figure}
    \centering
    \includegraphics[width=\linewidth]{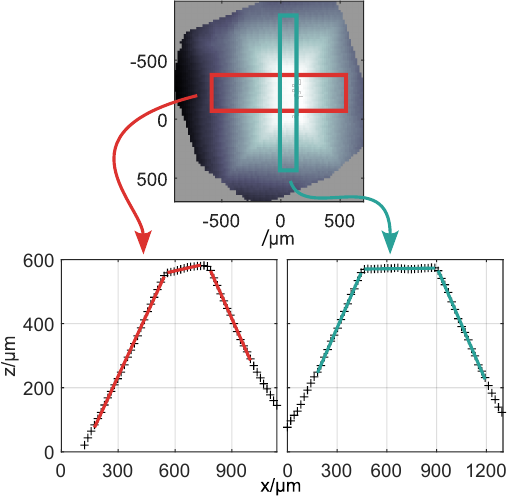}
    \caption{Height map of the reconstructed surface from figure 4 panel 4 with the horizontal and vertical slices used to measure the facet angles highlighted. The data for the slices is averaged with a linear regression performed for each facet, with the resulting gradients, plotted in color, used to calculate the angles. Figure adapted with permission from Radi\'{c} et al.\cite{Radic_3d_2024}. Reproduced from \textit{Appl. Phys. Lett.} 124, 204101 (2024), with the permission of AIP Publishing.}
    \label{fig:measurements}
\end{figure}

Figures \ref{fig:BigFigure}-\ref{fig:measurements} reproduces the first demonstration of a 3D topographic map produced by neutral atoms, as presented by Radi\'{c} et al.\cite{Radic_3d_2024}. Moreover, our 3D reconstruction method can be applied to a single detector instrument, which is particularly significant as most SHeM instruments presented in the literature only utilize a single detector. However, the requirement for image alignment introduces an uncertainty in all dimensions of a few pixels, further limiting reconstruction accuracy. Due to the micrograph field-of-view being non-circular, when they are superimposed through rotation during the reconstruction process some pixels are lost as they become under-constrained and must be discounted. Using figure \ref{fig:BigFigure} as an example,  24\% of the image area is unreconstructed, despite acquiring 5 micrographs. It should be possible to improve the ratio, and hence reduce the need for so many images, through more advanced scanning patterns and perhaps automated alignment during the acquisition process.

\section{Reconstruction using a multiple-detector instrument}
\label{sec:multi_detector}

Heliometric stereo on suitable multiple detector systems removes the need for rotational tracking entirely, but requires specialist hardware. The benefits of a multiple detector instrument are three-fold, 1) simultaneous imaging reduces imaging time by a factor of the number of detectors, 2) avoiding sample rotation increases reconstruction accuracy and further reduces imaging time because no sample alignment is required, and 3) decrease the likelihood that spatial drift in the sample or helium detection produces non-topographic contrast.  We use the purpose-built multiple detector 2nd generation SHeM at Cambridge to perform our experiments (B-SHeM)\cite{chenyang_bshem}. The B-SHeM has the capacity for up to four helium detectors whose collection apertures can be arbitrarily placed around the sample owing to the versatility of the instrument's defining optical element, the pinhole plate. The optical element can be 3D printed in plastic with complex internal geometries whilst retaining the (ultra)high-vacuum compatibility of traditionally machined metal using a simple baking method\cite{radicPlastics,Radic_3d_2024}.

\begin{figure}
    \centering
    \includegraphics[width=\linewidth]{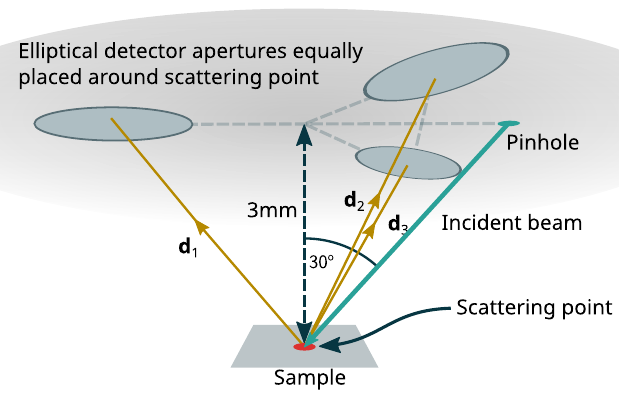}
    \caption{Schematic of the multiple detector B-SHeM instrument scattering geometry. A helium beam is incident at $\SI{30}{\degree}$ from the sample normal, and three detector apertures are placed around the scattering point in the same plane as the pinhole. The sample is $\SI{3}{\milli\metre}$ below the plane containing the pinhole and detector apertures.}
    \label{fig:bshem_instrument_schematic}
\end{figure}

In the presented work, we used the B-SHeM with three detectors (Hiden Analytical HAL/3F RC301 PIC30\cite{HidenAnalyticalSpecs}) to minimally constrain the reconstruction and provide a lower-bound on the method's accuracy. A schematic of the scattering geometry employed in the B-SHeM in the current work is shown in figure \ref{fig:bshem_instrument_schematic}. Three elliptical detector apertures are placed with equiangular spacing around the scattering point, each forming a circular region of solid angle when viewed from the scattering point.

\begin{figure*}
    \centering
    \includegraphics[width=\linewidth]{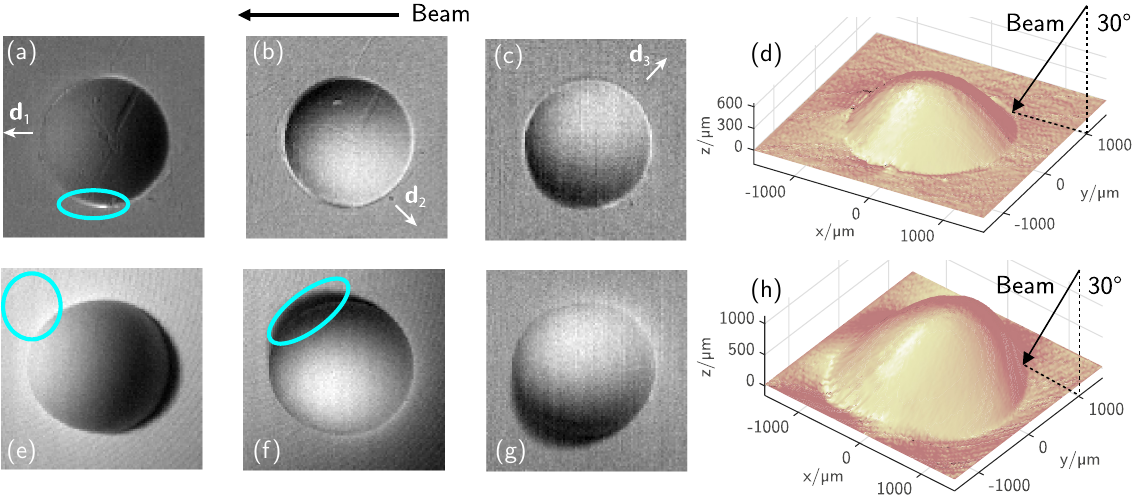}
    \caption{Micrographs, panels (a)-(c) and (e)-(g), and surface reconstructions, panels (d) and (h), of different sized hemispheres to test the effect of aspect ratio on reconstruction accuracy using a known sample geometry. Incident beam (right to left) and detection directions (d\textsubscript{x}) are labeled and consistent across all micrographs. Heliometric stereo reconstructions, using three Hiden Analytic mass spectrometers for detection \cite{HidenAnalyticalSpecs}, for spheres embedded into the substrate by different amounts, resulting in aspect ratios of, $0.36$ (d) and $0.52$ (h). The beam has a 30-degree incidence angle, as indicated in the top right corner of each surface reconstruction plot. The highlighted regions of the micrographs exhibit, multiple scattering resulting in a sharply defined bright region (a), masking resulting in a dark region of an image (e) and multiple scattering resulting in a broad bright region that displaces the featureless background (f). Micrographs acquired with $\SI{30}{\micro\metre}$ pixel size. \SI{500}{\micro\metre} scale bar included. Full imaging details in section SIII. Figure adapted from Zhao and Lambrick et al. \cite{chenyang_bshem}. C. Zhao and S.M. Lambrick et al., \textit{Vacuum}, Vol. 234, Article No. 114006, 2025; licensed under a Creative Commons Attribution (CC BY) license.}
    \label{fig:multi_det_3d}
\end{figure*}

To demonstrate multiple-detector heliometric stereo we selected a sample with known geometry for quantitative assessment of reconstruction quality. Two metal spheres with diameters $\sim\SI{2}{\milli\metre}$ were embedded by different amounts in a substrate and imaged in the B-SHeM to obtain three simultaneous images that were then used to generate a surface profile using the heliometric stereo method. Figure \ref{fig:multi_det_3d} shows the acquired SHeM micrographs and the corresponding reconstructions of the spheres. 

The reconstruction presented in this section marks a $\approx 5.4\times$  decrease in acquisition time per reconstructed pixel compared to a single detector implementation \cite{Radic_3d_2024}.  The decrease in acquisition time arises from both the simultaneous acquisition of images, and the complete usage of all pixels acquired, as noted in section \ref{sec:single-detector} 24\% of the image area is unreconstructed due to the different fields of view of the five micrographs. The single (A-SHeM) and multiple detector (B-SHeM) instruments have different detector dwell and response times. If all specific instrument parameters are ignored then the relative speed up would be $\approx \times 5 \times 1/(1-0.24)\approx \times6.6$ which accounts only for the simultaneous acquisition of images and the 24\% of pixels unused in the single detector reconstruction. A speed up of roughly $\sim 5-7\times$ is therefore achieved with the multiple detector approach similar to that presented here. We note that improvements could be made in the acquisition speed for single detector approach \textit{via} alternative, non-rectangular, raster patterns, which could reduce the fraction of the image area unreconstructed; these improvements may also allow for a reduction in the number of images required. In general, and for an ideal system the speed-up would be equal to the number of detectors used in the multiple detector setup --- $\times 3$ for the B-SHeM setup compared to an equivalent single detector setup with ideal image alignment --- but in practice this is a lower bound on the improvement because of mechanical drift and imperfect matching of raster patterns. Details of the imaging parameters for all images used for reconstructions is given in Supplementary Materials sections SII and SIII. In the current work, we also use a $\SI{30}{\degree}$ angle of incidence instead of normally incident beam in section \ref{sec:single-detector}. Non-normal incidence has significant implications for the reconstruction quality of vertical surfaces, which will be discussed in detail in section \ref{sec:masking_shadowing}.

Using the known spherical geometry and dimension of the two spheres we can evaluate the quality of the heliometric stereo reconstruction \textit{via} the overall root-mean-square (RMS) error, which here is defined as the squared deviation of the reconstruction from the known geometry averaged over all pixels and normalized by the known height of the sphere:
\begin{equation}
    \text{RMSE} = \frac{1}{H}\sqrt{\frac{\sum_\text{pix}(z_R - z_0)}{N_\text{pix}}}
\end{equation}
with $z_R$ being the reconstructed height and $z_0$ the height derived from the known geometry and $H$ the height of the sphere. The average is over all pixels considered to be part of the sample and not substrate, the areas used in the averaging are given in supplementary information SIV. The RMS gives a measure of the total deviation of the reconstruction from the true shape. Following from the definition used by Lambrick \& Salvador-Palau et al.\cite{Lambrick3d} the RMS shape error allows the reconstructed surface heights to vary: $z_2=\alpha z + \beta$ and we enforce the known maximum height of the sample, $H$. The parameters $\alpha$ and $\beta$ are found by finding the value of $z$ for the flat substrate, and the maximum height in the reconstruction, the difference between the maximum height and the flat substrate is then required to be equal to $H$. Note that because the scaling is based on a single known parameter, and not an optimization, the shape error is not restricted to being smaller than the RMS error, although that is often the case. The metric is useful as it separates two different errors in the reconstruction, that of the shape of the surface, and the overall height of the reconstruction. Where the shape is recovered reasonably well but the overall height is under- or over-estimated the shape error will come out smaller than the RMS error. If the shape itself is reconstructed poorly then the shape error will be comparable to the RMS error. Underestimation of the height was observed with some simulated data, when multiple scattering was included, by Lambrick \& Salvador-Palau et al.\cite{Lambrick3d} (figure 13) and therefore the shape error allows us to disentangle if the overall observed deviations are from this underestimation or due to false reconstructed shapes. Table \ref{tab:bshem_recon_errors} contains the calculated values of aspect ratios, defined here as the height of the sphere above the substrate over the diameter of the sphere, overall RMS errors and RMS shape errors for spheres A and B, alongside the signal-to-noise ratios (SNR) for each of the three detectors. We limit the calculation of the RMS and shape errors to the sample itself and exclude most of the flat substrate, details are given in the SI. 

The reconstruction of sphere A was qualitatively good with an overall RMS error calculated as 11.2\% which exceeds the 3.4\% predicted computationally by Lambrick \& Salvador-Palau et al.\cite{Lambrick3d} for non-normal incidence scattering conditions with 4 detectors. The difference can be attributed to two factors: (1) we use fewer detectors in the presented work than in the computational evaluation, 3 and 4, respectively. (2) The detectors used by B-SHeM have, on average, approximately a factor of 2 poorer SNR than the simulated detection where SNR=30 was set. Lambrick et al. found that the RMS error scales with the SNR. The precise scaling of the scaling of SNR and reconstruction accuracy is unclear, although the simulated efforts in figure 9 of \cite{Lambrick3d} seem to show that increasing the number of detectors has a similar effect to increasing the SNR of the images. Given the scaling in figure 9 of \cite{Lambrick3d} the difference in signal to noise between the simulated results and the current work is likely only a partial explainer of the difference in RMS error.

\begin{table}[h!]
\centering
\begin{tabular}{ccccccc}
\toprule
\multirow{2}{*}{Sample} &
  \multirow{2}{*}{\begin{tabular}[c]{@{}c@{}}Aspect \\ Ratio\end{tabular}} &
  \multicolumn{3}{c}{\begin{tabular}[c]{@{}c@{}}Signal-to-Noise Ratio\\ (SNR)\end{tabular}} &
  \multicolumn{2}{c}{\begin{tabular}[c]{@{}c@{}}RMS Error\\ (\%)\end{tabular}} \\ \cmidrule{3-7} 
         &      & Det. 2 & Det. 3 & Det. 4 & Overall & Shape \\ \midrule
Sphere A & 0.36 & 15.76  & 9.46   & 24.52  & 11.2\%                  & 9.9\%              \\
Sphere B & 0.52 & 17.20  & 14.37  & 33.28  & 13.8\%                  & 13.9\%             \\ \bottomrule
\end{tabular}
\caption{Signal-to-noise ratios and spatially averaged root-mean-square (RMS) (overall and shape) errors of the multiple detector reconstructions of spheres A and B from figure \ref{fig:multi_det_3d}. Spatially resolved errors are shown in figure \ref{fig:bshem_3d_errors} and further analyzed in section \ref{sec:masking_shadowing}.}
\label{tab:bshem_recon_errors}
\end{table}

By contrast, the reconstruction for sphere B is qualitatively poorer. Indeed, Table \ref{tab:bshem_recon_errors} shows that the scaling to produce the shape error is marginally larger than the raw RMS error, the difference is very small and we consider to be insignificant. The greater size of sphere B means that it exhibits greater masking and shadowing in its SHeM micrographs and consequently decreasing reconstruction accuracy \cite{Lambrick3d}. These effects are explored in more detail in the next section.

\section{The effects of masking, shadowing and multiple scattering}\label{sec:masking_shadowing}

\begin{figure}
    \centering
    \includegraphics[width=\linewidth]{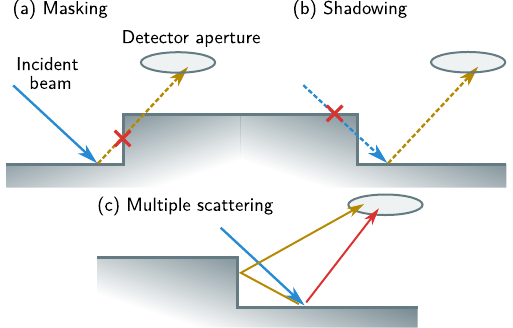}
    \caption{The mechanism of: (a) masking, where the line of sight between the scattering point and the detector aperture is blocked, rendering dark regions of micrographs. (b) shadowing, where the incident beam is blocked from regions of sample, resulting in those regions not appearing in micrographs. (c) multiple scattering, where atoms may scatter more than once to reach the detector, increasing the brightness of pixels in the helium micrograph.}
    \label{fig:masking_shadowing_multipleScattering}
\end{figure}

The effects of masking, shadowing, and multiple scattering in SHeM micrographs all affect the quality of a heliometric stereo reconstructions. We define each of these terms as the following\cite{Lambrick2018,Fahy2018,LambrickMultiple2020}, with the mechanisms being shown diagrammatically in figure \ref{fig:masking_shadowing_multipleScattering},

\begin{itemize}
\setlength\itemsep{-0.2 em}
    \item \textbf{masking - } an obstruction in the line-of-sight between the scattering site on the sample and the detector aperture,
    \item \textbf{shadowing - } an obstruction in the line-of-sight between the incident beam aperture and a spot on the sample,
    \item \textbf{multiple scattering - } any beam particle path that scatters more than once before detection.
\end{itemize}
It should be noted that the definitions of masking and shadowing used in the SHeM community are aligned with those used in optical, rather than electron, microscopy. An example is given by Linli Sun in their figure 2\cite{Sun2023} for the case of optical microscopy. The convention was adopted upon the first formalization of these contrast mechanisms in SHeM by Fahy et al. \cite{Fahy2018}.

As previously mentioned, these mechanisms break the key assumption of heliometric stereo: the direct relationship between scattered helium intensity and the surface orientation according to equation \eqref{eq:cosine}. 

Masking, highlighted in figure \ref{fig:multi_det_3d} (f), causes dark regions in images, which will be reconstructed (equation \eqref{eq:ch3D:solvePhotostereo2}) as regions parallel to the detection direction. Shadowing results in areas of the sample where no information is gathered. Multiple scattering results in brighter regions of images, generally near vertical walls, recesses or similar features, such as in figure \ref{fig:multi_det_3d} (a), where a small recess at the edge of the embedded sphere results in enhanced contrast\cite{LambrickMultiple2020}, and the elevated brightness on the substrate highlighted in figure \ref{fig:multi_det_3d} (e) as well as in smaller amounts on the sample itself. Bright regions caused by multiple scattering will also distort the solution to equation \eqref{eq:ch3D:solvePhotostereo2}. 

We use the two spheres shown in figure \ref{fig:multi_det_3d} as our quantitative test of the impacts of masking and shadowing. The spheres provide a good test sample to quantify the effect of masking and shadowing on the method because both spheres have identical, and known, geometries, aside from sphere B having a larger aspect ratio because it is embedded in the substrate to a lesser degree. With masking and shadowing being geometrically defined as lines-of-sight between the sample and the detector or the incident aperture, respectively, when the aspect ratio of a feature increases the degree of masking and shadowing must also increase. This investigation is similar to that performed on simulated data in our previous work, where reconstruction accuracy was plotted as a function of the aspect ratio of features\cite{Lambrick3d}. 

\begin{figure}
    \centering
    \includegraphics[width=0.9\linewidth]{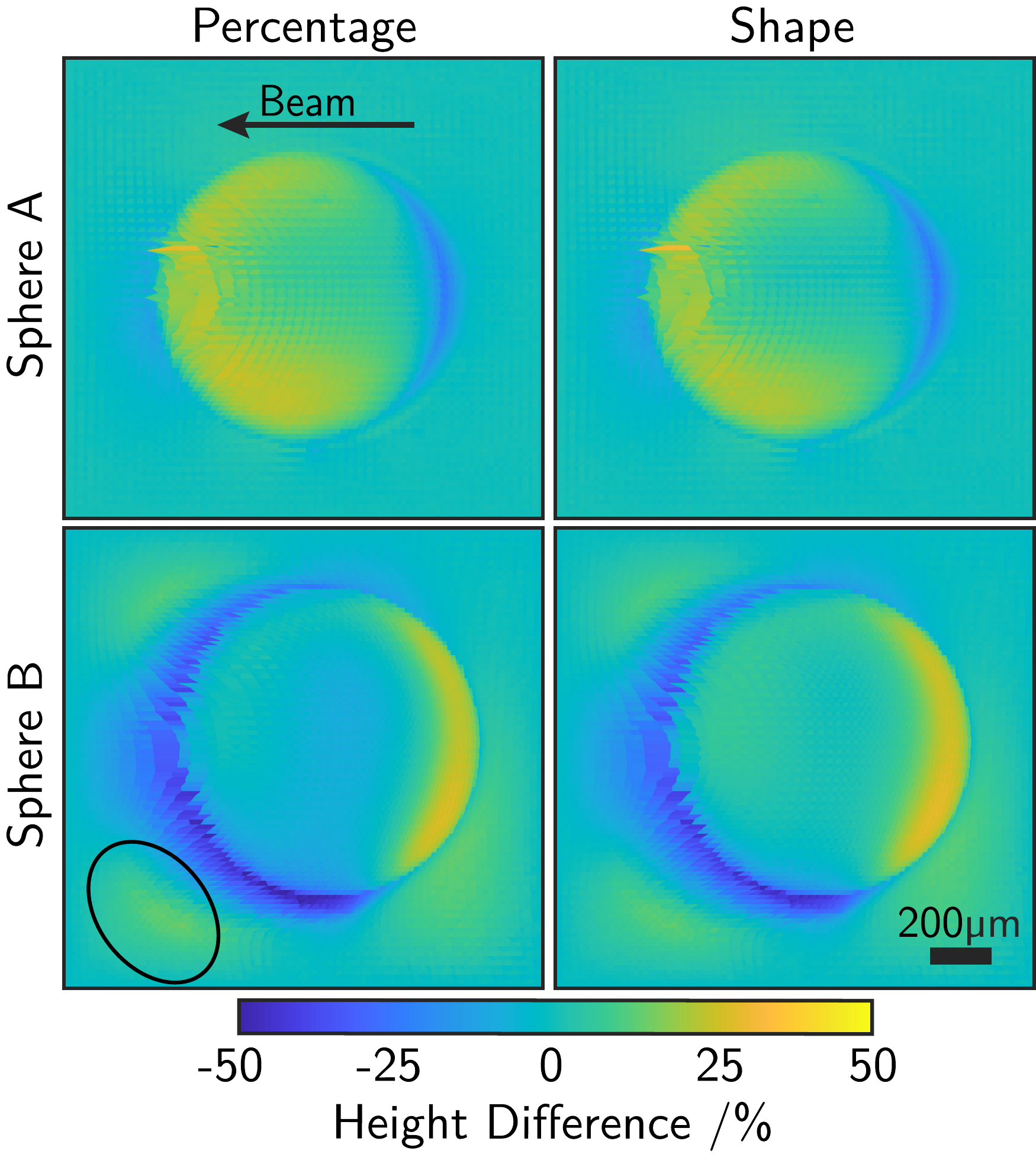}
    \caption{Plots of percentage height difference and shape difference for spheres A and B, yellow regions lie higher than the expected surface, and blue lower. The overall RMS error and RMS shape error are calculated by averaging the differences. The shape difference excludes the contribution from an error in the overall height of the spheres. The RMS for total error and shape error were 11.2\% and 9.9\% for sphere A, as well as 13.8\% and 13.9\% for sphere B, respectively. The incident helium beam propagates from right to left. The highlighted region shows an area of substrate with higher than actual height reconstruction due to multiple scattering.}
    \label{fig:bshem_3d_errors}
\end{figure}

Figure \ref{fig:bshem_3d_errors} uses the known geometry of the spheres to plot the difference in reconstructed height, and the difference in reconstructed shape, from the expected geometry. Yellow regions lie higher than the expected surface, and conversely blue regions lie lower, green regions lie at the expected height. 

For the lower aspect ratio sphere A, there is reasonably good accuracy across the sample, with only some small regions of poorer precision. In particular, the accuracy drops near the edge of the sphere, likely due to a crevice where the sphere is embedded into the surface, causing masking and multiple scattering. These errors are, however, locally confined, and do not cause errors away from the edges, similar to results found with simulated data\cite{Lambrick3d}. Across the sphere itself the errors are asymmetric and match with the direction of the beam, but remain generally modest, generally reaching no more than double the average error. The reconstruction also marginally underestimates the overall height of the sphere, recovering it at 94\% of the expected height.

We attribute the high quality reconstruction to the fact that the basic assumptions of heliometric stereo are held across almost the entire sample area: sphere A has a relatively low aspect ratio and no vertical faces, thus avoiding masking and exhibiting only relatively modest multiple scattering in the helium micrographs.

By contrast, while sphere B has the same basic shape as sphere A its increased aspect ratio results in more masking and shadowing. As can be seen in figure \ref{fig:bshem_3d_errors} the reconstructed surface lies noticeably higher towards the incident beam, and lower away from it, resulting in a visibly deformed reconstruction (figure \ref{fig:multi_det_3d}). If we inspect quantitatively, we also find that the magnitude of the maximum deviation from the expected surface is also significantly greater than for sphere A, reaching $[-\SI{52,3}{},+\SI{28.3}{}]\%$ and $[-\SI{48.4}{},+\SI{29.8}{}]\%$ in percentage and shape, respectively. The greatest errors occur near the edge of the sphere, where there is non-negligible masking in all three micrographs, one of which is highlighted in figure \ref{fig:multi_det_3d} (f), and where the surface of the sphere becomes approximately parallel to the beam. In the RMS errors in table \ref{tab:bshem_recon_errors} we also see that for sphere B, where masking is significant the error does not decrease when the reconstruction is allowed to scale (shape error), unlike for sphere A. A reduction in the error when scaling is allowed suggests that for sphere A the shape of the sample was well recovered, but that the height was underestimated. The lack of change (or negligible increase) in RMS error when the reconstruction is scaled suggest that there are significant errors in the shape of the sample, not merely a overall underestimation of the height. In the simulated results of Lambrick \& Salvador Palau et al.\cite{Lambrick3d} (figures 13/14) it was found that inclusion of (modest) multiple scattering induced an underestimation of the overall height, but that the shape was still recovered well; however, when the aspect ratio of samples was increased and masking appeared the shape was effected strongly. We believe the differences between the RMS and shape error in spheres A/B are due to the absence/presence of significant regions of masking, which sphere A does not posses and sphere B does. The simulated results suggest that by excluding masked regions from reconstruction more accurate shapes can be recovered, however, in order to exclude pixels and still perform a reconstruction more than 3 images are needed, which was not possible with the current experimental setup.

\begin{figure}
    \centering
    \includegraphics[width=\linewidth]{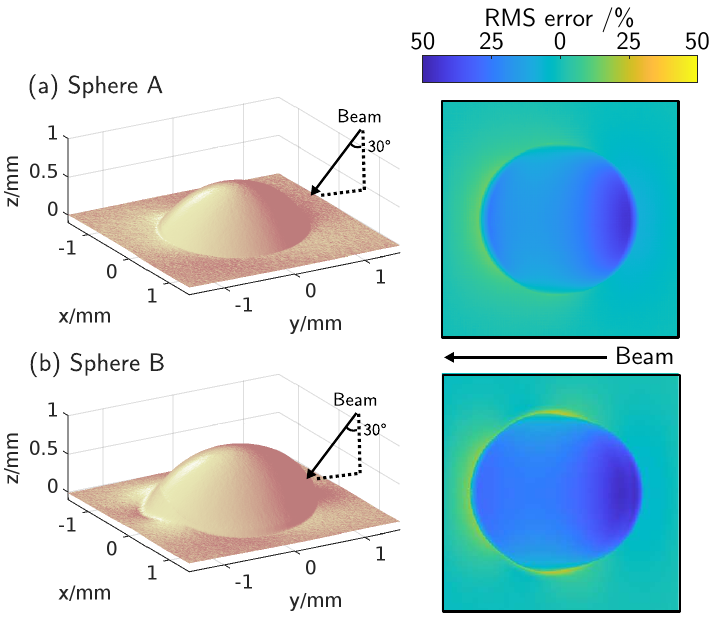}
    \caption{Heliometric stereo reconstructions and percentage error maps using simulated ray tracing data for samples of the same topography as sphere A and B. We note that both simulated reconstructions underestimate the overall height, and this underestimation is the main error in both cases; whereas for the experimental data the heights are recovered better but other errors are present, especially in the case of sphere B. RMS error values for the whole area of the reconstruction are 11\% and 15\% for sphere A and B respectively.}
    \label{fig:simulated_results}
\end{figure}

We find for both spheres that the overall height was recovered well, with 94\% and 109\% of the expected height recovered for spheres A and B respectively. This performance is slightly better than expected from simulations, which generally demonstrated a more significant underestimation of the overall height than we observe experimentally. Figure \ref{fig:simulated_results} gives the reconstructed surfaces and percentage error maps (same scale as figure \ref{fig:bshem_3d_errors}). Here, as with the experimental data, we see a good quality reconstruction for sphere A, with the shape well preserved, but some distortion of the shape with sphere B. Unlike for the experimental data we underestimate the height of the sphere, recovering 85\% and 76\% of the expected height, as can be seen in the error maps. These values align with previous results where the dependence of the error on aspect ratio was explored\cite{Lambrick3d}. The qualitatively different error maps between the simulated and the experimental data are interesting, and suggest that the ray-tracing simulations, which have a good track record of producing representative SHeM data\cite{Lambrick2018,LambrickDiffuse2022,lifPaper,LambrickMultiple2020} may be incomplete for the current experimental setup.

\begin{figure}
    \centering
    \includegraphics[width=\linewidth]{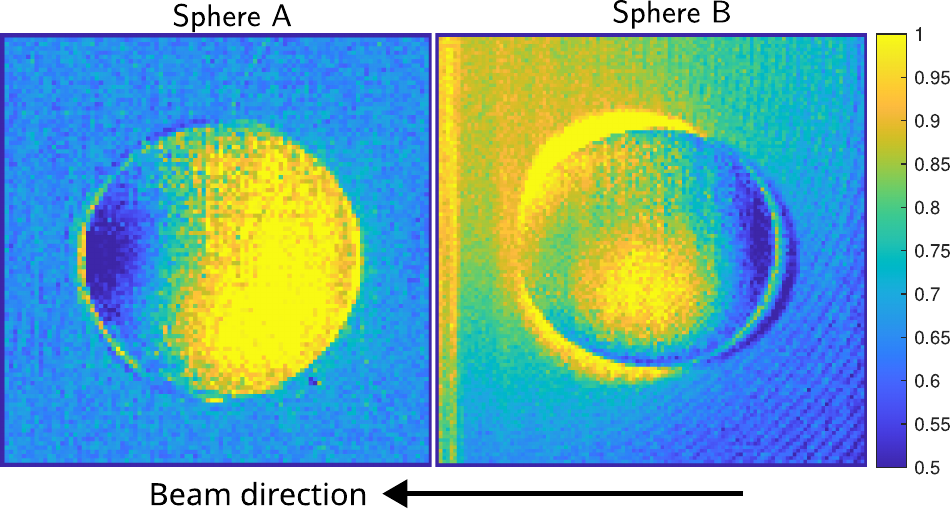}
    \caption{Albedo factor calculated during the heliometric stereo reconstruction for sphere's A and B. Generally the variation is slow, but rapid changes occur near masking features which break the assumption of a relationship between intensity and topography. We also observe variation across the reconstruction for sphere B, which links to the varying brightness across the micrographs used in the reconstruction -- see figure \ref{fig:multi_det_3d} (e)-(g). The range of albedo factors observed is similar to that seen with the simulated data in figure \ref{fig:simulated_results}, with the majority of pixels falling between $\SIrange{0.6}{1.0}{}$ with some localized regions of outliers.}
    \label{fig:albedo}
\end{figure}

The reconstruction process outlined by equations \ref{eq:ch3D:basic_photostereo} allows the recovery of the `albedo factor'. While we do not attribute direct physical meaning to the recovered value we do observe rapid changes in the recovered value where deviant contrast features such as masking and strong multiple scattering arise. Figure \ref{fig:albedo} maps the recovered albedo factor for sphere's A and B. In most areas the albedo varies slowly, however, strong changes occur near masking in sphere B, where we know that the $I\sim\cos\theta$ relationship breaks down. For the test sample these maps of the albedo do not directly inform us of anything new, however, for more complicated samples where interpretation of the images is not as clear and the sample topography not known the albedo may become helpful in identifying regions of low accuracy reconstruction.


Sphere B also demonstrates the effects of multiple scattering. On the flat substrate near the sphere there is a slight distortion in the reconstruction, the surface is higher than expected, \emph{e.g.} in highlighted region in figure \ref{fig:bshem_3d_errors}, which is likely to arise from brightening of the substrate, such as in figure \ref{fig:multi_det_3d} (e). It is possible the multiple scattering has also contributed to the reduced height of the final reconstruction. Here, the intensity of the helium micrographs has been increased in a similar manner to that shown in figure \ref{fig:masking_shadowing_multipleScattering} (c), causing distortions in the reconstruction.

As, in most circumstances, multiple scattering only accounts for a small fraction of SHeM contrast these distortions have remained modest. There are specific sample geometries where the effects of multiple scattering are severe and unavoidable across much of the micrograph, such as high aspect ratio holes or troughs into a surface, however in most reported micrographs strong regions of multiple scattering tend to be localized to only part of the sample. An example of the obfuscation of the bottom of a narrow trough solely due to multiple scattering is shown by Lambrick et al.\cite{LambrickMultiple2020} in Figure 1. One can liken the outcome to the tip convolution problem in AFM and STM. Multiple scattering is intrinsic to neutral atom microscopy and can only be controlled to a limited degree with instrument geometry\cite{Fahy2018}, thus future applications of heliometric-stereo will need to consider it carefully. The instrument geometry is especially relevant in pinhole microscopes designed to operate at short working distances, such as that presented by Witham et al.\cite{Witham2011}, where multiple scattering from the pinhole plate can become significant.

Shadowing, where the line-of-sight between incident beam and sample is blocked, is not present to a significant degree in the test samples considered here, however, it could lead to problems in reconstructions as parts of the sample could be missed. For a normal incidence configuration, such as that used in section \ref{sec:single-detector}, shadowing will occur at overhangs and vertical faces. For a non-normal incidence configuration, such as used in section \ref{sec:multi_detector}, vertical faces and overhangs can be captured if they face the beam, but regions facing away from the beam will be hidden. One approach to minimize the effect of shadowing would be to combine multiple reconstructions from different incidence conditions with the sample in different orientations, however, doing so would increase acquisition time.

Masking, which is intrinsic to neutral atom imaging, can be partially mitigated by changing the relative positions of the beam and the detectors: the larger the angular distance between the detector and beam, the greater the amount of masking\cite{lambrick_2021}. Both implementations in the current work already attempt to keep the angular distance between detector and beam modest, although it was found that the desire to reduce the angular distance conflicts with the limited space around the sample in both single and multiple detector SHeM, and in practice some compromise had, and will have, to be made in the design. For future instruments or modifications intended for the use of heliometric stereo, attempts to reduce masking must be kept in consideration alongside other requirements of the instrument.

All of these changes can, in principle, be implemented onto current single and multiple detector SHeM instruments cheaply and rapidly without significant hardware changes because the illumination and detection geometry is solely dictated by the pinhole plate optical element.

\begin{figure}
    \centering
    \includegraphics[width=\linewidth]{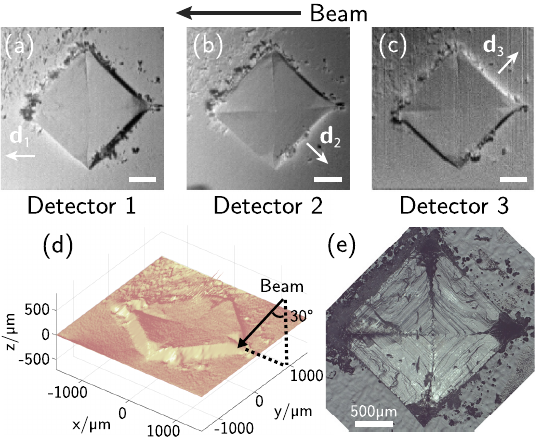}
    \caption{A heliometric stereo reconstruction of a salt crystal on a flat substrate. The three micrographs in panels (a)-(c) were used to generate the minimally constrained reconstruction in panel (d). An optical image is included in panel (e) for qualitative comparison. The salt crystal edges exhibit both masking and multiple scattering here, however, as the area is limited the overall reconstruction is still qualitatively sound when compared to an optical image in the lower right. Micrographs acquired with $\SI{20}{\micro\metre}$ pixel size. \SI{500}{\micro\metre} scale bar included. Full imaging details in section SIII. Figure adapted from Zhao and Lambrick et al. \cite{chenyang_bshem}. C. Zhao and S.M. Lambrick et al., \textit{Vacuum}, Vol. 234, Article No. 114006, 2025; licensed under a Creative Commons Attribution (CC BY) license.}
    \label{fig:salt}
\end{figure}

We include a reconstruction of a salt (sodium chloride) crystal in figure \ref{fig:salt} using the same experimental setup as section \ref{sec:multi_detector}. Panels (a)-(c) show the helium micrographs used, panel (d) the resulting reconstruction and panel (e) an optical micrograph of the sample. The crystal has relatively low aspect ratio overall, but does posses a few vertical features at its edges that create strong shadowing or masking. Despite the local regions of shadowing and masking a reconstruction is generated that captures the overall topography of the sample well, highlighting the robustness of our method. By comparison to an optical microscope image (Keyence VHX-7000 microscope) we see that the heliometric stereo reconstruction reproduces the fine faceted structure on the surface of the crystal, including the cracks connecting the corners of the crystal, along with very fine ripples that appear parallel to the sides of the crystal forming concentric squares, which are of a size comparable to that of a single pixel in the SHeM micrographs. We note that these very fine features would not be recoverable with a single detector instrument as the lateral accuracy would be degraded by the image alignment procedure.

We have demonstrated that heliometric stereo can be applied successfully both with a single detector SHeM, section \ref{sec:single-detector}, and with a multiple detector SHeM, section \ref{sec:multi_detector}. The primary advantage explored thus far for the multiple detector approach is the speed up in measurement time, with a speed up of $\times 5.4$ directly achieved in the presented results, and speed ups of roughly $\times 5-7$ achievable with the presented setups depending on specific parameters used. An apparent drawback of the multiple detector approach presented here, where a non-normal incidence angle is used, is the limited number of micrographs available, 3 instead of 5. In general more micrographs increase the SNR, but more importantly, they can be used to get around masking issues\cite{Lambrick3d} by allowing certain data regions to be excluded from the reconstruction while still having sufficient data to solve equations \ref{eq:ch3D:basic_photostereo}. In the multiple detector case any incidence angle can be used, while the single detector case is limited to normal incidence. As the reconstruction method performs poorly for samples areas parallel to the beam, varying the incidence angle for the specific sample of interest could be a benefit for some systems. Alternatively if a normal incidence beam were to be used for a multiple detector system a hybrid approach can be employed, where rotations are used to multiply the number of detector, \emph{e.g.} a single rotation could provide 6 effective detectors with a 3 detector microscope setup. The hybrid approach would yield more modest speed ups, but would provide more data points and hence likely a high reconstruction quality than either approach presented in the current work.

As well as developments of the experimental hardware to overcome the limitations imposed by masking and multiple scattering, we can also envisage future improvements to the computational methods. For example, stereo-photogrammetry or masks in micrographs can be used to give accurate dimensions for specific features\cite{Myles2019,LambrickMultiple2020}, these could be used to calibrate and improve the accuracy of heliometric-stereo reconstructions where height of features can be underestimated (such as the larger sphere in figure \ref{fig:multi_det_3d}). In addition, more involved iterative methods could be used to reduce the uncertainties arising from multiple scattering, an issue that may become even more relevant if microscopes move to shorter working distances to improve spatial resolution. By using tools developed to predict contrast in SHeM that account for multiple scattering\cite{Lambrick2018} the multiple scattering intensities from the initially reconstructed surface could be predicted, those predictions could then be used to inform/correct a second (or further) reconstruction(s). Potentially more advanced parametric, or even non-parametric\cite{zickler_helmholtz_2002}, models for the BRDF could be used, loosening the hard assumption of diffuse cosine-like scattering, however, adding more parameters would entail the collection of a greater number of micrographs per sample.

\section{Conclusion}

We present accurate surface profile reconstructions using both a single detector and a multiple detector instruments. Best results were obtained with samples suited to the technique: those that conform to the key assumptions of the method, most notably the assumption that the topography can be described by a single valued function $z = f(x,y)$. 
We expand upon the adaptation of the heliometric stereo method, as reported for a single detector instrument by Radi\'{c} et al.\cite{Radic_3d_2024}, by presenting results to demonstrate the effect of excessive masking and shadowing due to samples with vertical edges and high aspect ratios. We suggest instrumentation changes to the illumination and detection geometry that can mitigate masking and shadowing, to improve reconstruction quality, that can be easily implemented into current SHeM instruments. We find, by comparing two different aspect ratio partial spheres, similarly to previous simulated results that reconstructions work well provided  aspect ratios are kept $\leq 1$. 

Furthermore, we present a realistic sample of unknown geometry, the salt crystal in figure \ref{fig:salt}, and verify against optical microscopy that heliometric stereo successfully reproduces fine structures on the order of the $\SI{20}{\micro\metre}$ pixel size. 
It was found that even when the assumptions were broken locally, \emph{e.g.} by overhangs and vertical faces with the salt crystal sample, (features that prove challenging in many other techniques too)  only local errors were introduced to the reconstructions.

Comparing the single and multiple detector reconstructions we highlight inherent advantages to the multiple detector implementation, including significantly faster imaging time, minimal manual intervention, and the recovering of pixel-level topographic information. With previously reported lateral SHeM spot sizes down to $\SI{315}{nm}$\cite{Witham2014}, and recent advances in neutral beam generation\cite{skimmer_paper}, existing SHeM instruments could, with minor modification, access sub-optical resolution surface reconstructions for moderate to high aspect ratio samples made of soft, insulating, optical transparent or electrically sensitive materials.

The success of the reconstruction process further validates recent work demonstrating that diffuse scattering is dominant for technological surfaces in SHeM. Looking ahead, we anticipate that the methods presented by Myles et al.\cite{Myles2019} or the use of contrast features as presented by Lambrick et al.\cite{LambrickMultiple2020} will be combined with heliometric stereo for higher accuracy surface metrology, especially as SHeM resolution pushes into the nanoscale.

\section*{Supplementary Material}
See the supplementary material for additional information on the sample used for single detector reconstruction in Section \ref{sec:single-detector}, detailed SHeM imaging parameters used to acquire micrographs in figures \ref{fig:BigFigure}, \ref{fig:multi_det_3d}, \ref{fig:salt}, and further details on the RMS error calculation.

\section*{Supplementary Material}

\preprint{AIP/123-QED}

\title[Heliometric stereo: a new frontier in surface profilometry]{Heliometric stereo: a new frontier in surface profilometry}


\author{\underline{Aleksandar Radi\'{c}}}
\email[Author to whom correspondence should be addressed: ]{ar2071@cam.ac.uk}
\affiliation{Department of Physics, Cavendish Laboratory, 19 J.J. Thomson Avenue, University of Cambridge, Cambridge, CB3 0HE, UK}

\author{\underline{Sam M. Lambrick}}
\affiliation{Department of Physics, Cavendish Laboratory, 19 J.J. Thomson Avenue, University of Cambridge, Cambridge, CB3 0HE, UK}
\affiliation{ 
Ionoptika Ltd, Units B5-B6, Millbrook Close, Chandlers Ford, Southampton, S053 4BZ, UK}

\author{Chenyang Zhao}
\affiliation{Department of Physics, Cavendish Laboratory, 19 J.J. Thomson Avenue, University of Cambridge, Cambridge, CB3 0HE, UK}

\author{Nick A. von Jeinsen}
\affiliation{Department of Physics, Cavendish Laboratory, 19 J.J. Thomson Avenue, University of Cambridge, Cambridge, CB3 0HE, UK}

\author{Andrew P. Jardine}
\affiliation{Department of Physics, Cavendish Laboratory, 19 J.J. Thomson Avenue, University of Cambridge, Cambridge, CB3 0HE, UK}

\author{David J. Ward}
\affiliation{Department of Physics, Cavendish Laboratory, 19 J.J. Thomson Avenue, University of Cambridge, Cambridge, CB3 0HE, UK}
\affiliation{ 
Ionoptika Ltd, Units B5-B6, Millbrook Close, Chandlers Ford, Southampton, S053 4BZ, UK}

\author{Paul C. Dastoor}
\affiliation{Department of Physics, Cavendish Laboratory, 19 J.J. Thomson Avenue, University of Cambridge, Cambridge, CB3 0HE, UK}
\affiliation{Centre for Organic Electronics, Physics Building, University of Newcastle, Callaghan, NSW 2308, Australia}
\date{\today}

\maketitle
\end{comment}   
\setcounter{figure}{0}
\renewcommand{\figurename}{Fig.}
\renewcommand{\thefigure}{S\arabic{figure}}

\setcounter{table}{0}
\renewcommand{\tablename}{TABLE}
\renewcommand{\thetable}{S\arabic{table}}

\setcounter{section}{0}
\renewcommand{\thesection}{S\Roman{section}}

\section{Single detector sample SEM image}
\label{sec:big_tetra_sem}
\begin{figure}[h]
    \centering
    \includegraphics[width=0.95\linewidth]{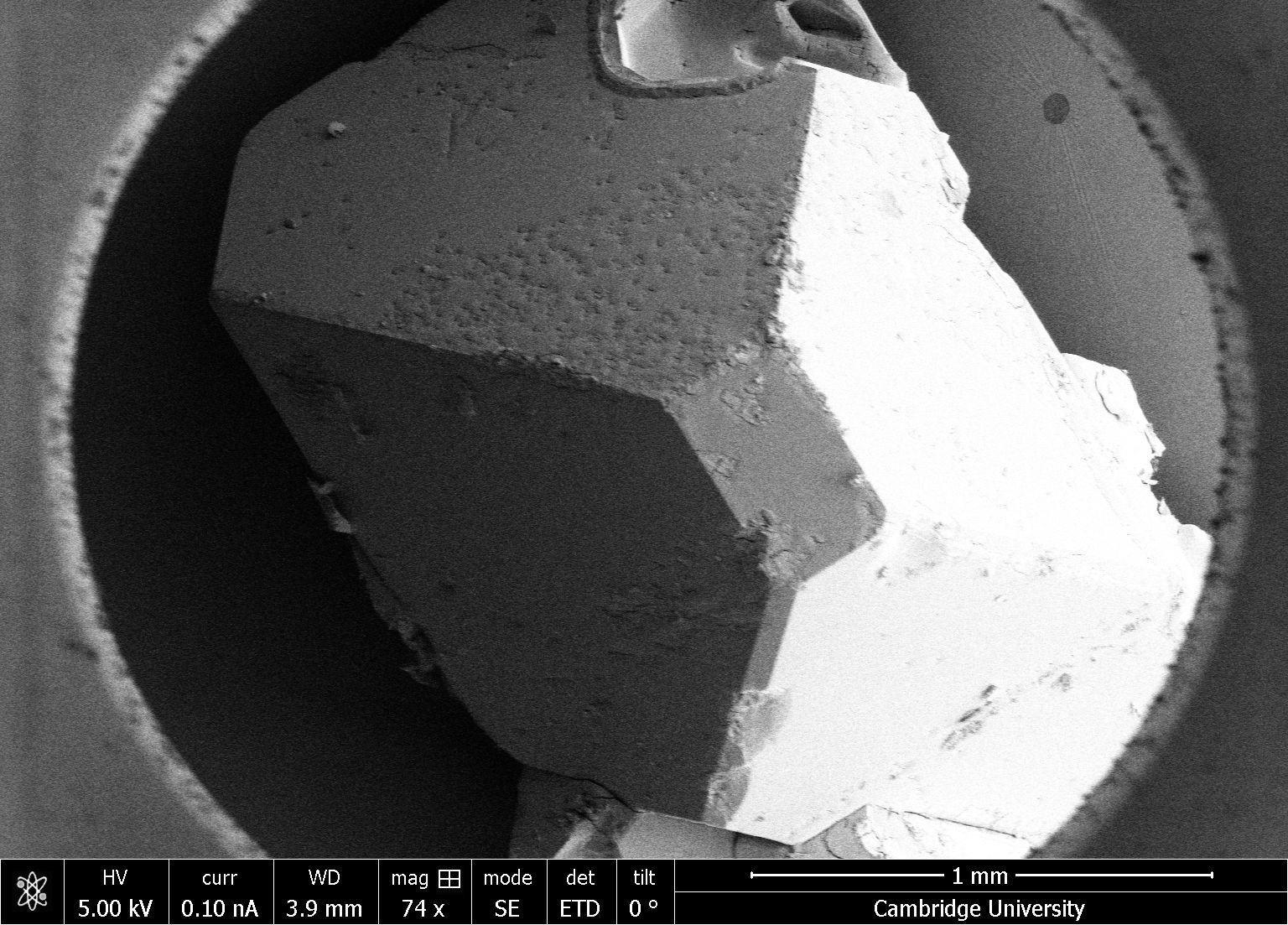}
    \caption{SEM image of the gold coated aluminum potassium sulfate ($KAl(SO_{4})_{2}. 12 H_{2}O$) crystal reconstructed in figure 4 taken prior to SHeM imaging.}
    \label{fig:big_tetra_sem}
\end{figure}

\section{Imaging parameters - single detector}
\label{sec:image_params_single}

The total acquisition time for the micrographs in figure 4 was 14 hours and 53 minutes (2x images with acquisition time of 3 hours 25 minutes and 3x images with acquisition time of 2 hours 41 minutes). Pixel sizes of $\SI{20}{\micro\metre}$ were used with a dwell time of $\SI{750}{\milli\second}$ for 2x images and $\SI{495}{\milli\second}$ for 3x images. $101\times101$ pixels were used for the micrograph.

\section{Imaging parameters - multiple detector}
\label{sec:image_params_multi}

The acquisition time for all three micrographs needed for a single reconstruction of a sphere, as shown in figure 7 (same parameters used for both spheres), was 5 hours and 14 minutes for images with dimensions $121\times121$ pixels and dwell time $\SI{1000}{\milli\second}$ per pixel.

The acquisition time for all three micrographs needed for the reconstruction of a salt crystal in figure 10 was 8 hours and 8 minutes for images with dimensions $151\times151$ pixels and dwell time $\SI{1000}{\milli\second}$ per pixel.




\section{Error calculation areas}

\begin{figure}[h]
    \centering
    \includegraphics[width=\linewidth]{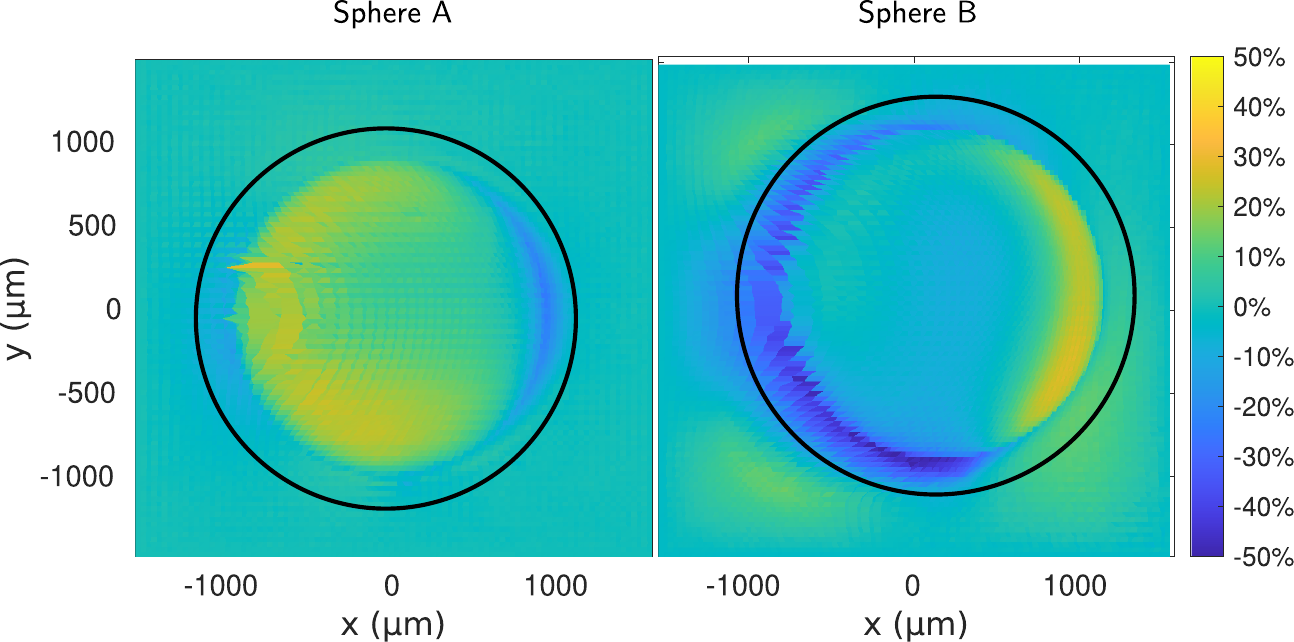}
    \caption{Percentage error plots of the reconstructions of sphere A and B (same data as in figure 9) overlaid with circles representing the regions where the error was calculated over. Area outside the circle was considered substrate and neglected during the calculation.}
    \label{fig:error_regions}
\end{figure}

For the calculation of the RMS and shape error most of the flat substrate was not included in calculations. A circular region centered on the sphere was used, these circles are drawn over the percentage error in figure \ref{fig:error_regions}.

\section*{Acknowledgments} 
The work was supported by EPSRC grant EP/R008272/1. The authors acknowledge support by the Cambridge Atom Scattering Centre (\url{https://atomscattering.phy.cam.ac.uk}) and EPSRC award EP/T00634X/1. The work was performed in part at \textsc{corde}, the Collaborative R\&D Environment established to provide access to physics related facilities at the Cavendish Laboratory, University of Cambridge. SML acknowledges support from IAA award EP/X525686/1 and funding from Mathworks Ltd. We would like thank Boyao Liu for useful discussions. The authors acknowledge support from Ionoptika Ltd.

\section*{Author declarations}
\subsection*{Conflict of interest}
The authors declare the following financial interests/personal relationships which may be considered as potential competing interests: Sam Lambrick reports a relationship with lonoptika Ltd. that includes: consulting or advisory. David Ward reports a relationship with lonoptika Ltd. that includes: consulting or advisory. The other authors declare that they have no known competing financial interests or personal relationships that could have appeared to influence the work reported in this paper.

\section*{Data availability} Supporting data packs containing both raw and processed data are available. Data for results with the Cambridge A-SHeM (figures \ref{fig:BigFigure} and \ref{fig:measurements}) can be found at \url{http://doi.org/10.17863/CAM.107537} . Data for figure \ref{fig:salt} (and some of the data for figure \ref{fig:multi_det_3d}) can be found at\url{https://doi.org/10.17863/CAM.114465}. A data pack for figures \ref{fig:multi_det_3d}, \ref{fig:bshem_3d_errors}, \ref{fig:simulated_results}, \ref{fig:albedo}, and the optical data in figure \ref{fig:salt} can be found at \url{https://doi.org/10.5281/zenodo.15548542}.

\section*{References}
\bibliographystyle{apsrev4-2}
\bibliography{heliometric_references}

\end{document}